\documentclass[twocolumn]{revtex4}
\def\address{\affiliation}

\usepackage[dvips]{graphicx}
\begin{document}

\title
[
Out-of-plane dielectric constant of
Bi$_2$Sr$_2$Dy$_{1-x}$Er$_x$Cu$_2$O$_8$
]
{
Out-of-plane dielectric constant and
insulator-superconductor transition in 
Bi$_2$Sr$_2$Dy$_{1-x}$Er$_x$Cu$_2$O$_8$ single crystals
}

\author{
T Takayanagi, M Kogure and I Terasaki 
}

\address{
Department of Applied Physics, Waseda University,
Tokyo 169-8555, JAPAN
}

\date{\today}

%--------------------------------------------------
%     Abstract
%--------------------------------------------------

\begin{abstract}
The out-of-plane dielectric constant of the parent insulator
of the high-temperature superconductor
Bi$_2$Sr$_2$Dy$_{1-x}$Er$_x$Cu$_2$O$_8$
was measured and analysed from 80 to 300 K 
in the frequency range of $10^6$-$10^9$ Hz.
All the samples were found to show a fairly large value of 10-60, 
implying some kind of charge inhomogeneity in the CuO$_2$ plane.
Considering that the superconducting sample 
Bi$_2$Sr$_2$Ca$_{0.75}$Pr$_{0.25}$Cu$_2$O$_8$ also shows 
a similar dielectric constant, 
the charge inhomogeneity plays an important role
in the insulator-superconductor transition.
\end{abstract}

\maketitle  % for ReVTeX4

%--------------------------------------------------
%     Introduction
%--------------------------------------------------

\section{Introduction}
A cuprate with a nearly undoped CuO$_2$ plane shows high resistivity 
and antiferromagnetic order at low temperature, 
which is called a parent insulator of high-temperature superconductors (HTSC).
With doping, an insulator-metal transition (IMT) arises, 
and the system changes from the parent insulator to HTSC.
For IMT, the dielectric constant $\varepsilon$ is of great importance 
in the sense that it provides a measure of localisation length in the insulator. 
In particular, $\varepsilon$ for doped Si 
divergingly increases with a critical exponent toward the IMT \cite{castner}.

Chen {\it et al.} first pointed out 
the importance of $\varepsilon$ for the IMT of HTSC \cite{Chen}. 
They found that $\varepsilon$ of La$_2$CuO$_{4+\delta}$
diverged only along the in-plane direction toward the IMT,
and stated that the IMT of HTSC was two-dimensional.
However, their study was done only for La$_2$CuO$_{4+\delta}$, 
in which a quasi-static stripe fluctuates in the CuO$_2$ plane \cite{stripe}.
Since the stripe fluctuation is a kind of charge modulation and/or
a charge disproportion,
it might cause a large dielectric response.
Perhaps related to this, (La,Sr)$_2$CuO$_{4+\delta}$ shows 
structure phase transition \cite{takagi}, oxygen ordering \cite{inaguma} and 
phase separation \cite{cho},
which might also give a large dielectric response. 
Thus it is necessary to study another class of HTSC, 
which is far away from the stripe- or lattice-instability.

We think that Bi$_2$Sr$_2$RCu$_2$O$_8$ (Bi-2212) is the best candidate.
This class of HTSC shows no structural phase transition below 300 K,
and is expected to be away from the stripe instability.
(The only exception is the Zn-doped Bi2212 of Ref.\cite{akoshima}.)
We should also note that Bi-2212 has only one equivalent Cu site, 
unlike the CuO chain in YBa$_2$Cu$_3$O$_7$.
We have studied the charge transport of the parent insulator 
of Bi-2212 \cite{Kitajima,terra,Takemura}, and have found that 
their physics is as rich as the physics of HTSC themselves.
In this paper, we report on measurement and analysis of 
the dielectric constant of Bi$_2$Sr$_2$Er$_{1-x}$Dy$_x$Cu$_2$O$_8$
single crystals along the out-of-plane direction.

%--------------------------------------------------
%     Experimental
%--------------------------------------------------
\section{Experimental}
Single crystals of Bi-2212 were grown by self-flux method.
The growth conditions and the sample characterisation were 
described in Ref.\cite{Kitajima}. 
The prepared crystals of 
Bi$_2$Sr$_2$Dy$_{1-x}$Er$_x$Cu$_2$O$_8$ 
($x$=0, 0.3, 0.5, 0.7 and 1.0) were insulating, 
and the doping levels of the as-grown crystals were 
slightly different for different $x$. 
An energy-dispersive x-ray (EDX) analysis revealed 
that Sr was slightly excessive, 
and the content was estimated to be 
Bi$_{2.2}$Sr$_{3.4-y}$R$_y$Cu$_2$O$_{8+\delta}$ 
(within an experimental error of 10\%). 
The R content $y$ was 1.0 for R=Dy and 0.7 for R=Er. 
Thus a real composition would be something like 
Bi$_2$Sr$_2$DyCu$_2$O$_8$
and Bi$_2$Sr$_2$(Sr$_{0.3}$Er$_{0.7}$)Cu$_2$O$_8$
for R= Dy and Er, respectively. 
Thus a solid solution between Er and Dy gives
Bi$_2$Sr$_2$Dy$_{1-x}$(Sr$_{0.3}$Er$_{0.7}$)$_{x}$Cu$_2$O$_8$,
which enables us to finely tune the doping level 
from highly insulating ($x$=0) to slightly doped level ($x$=1). 
Instead of the real composition,
we will call the prepared samples
Bi$_2$Sr$_2$Dy$_{1-x}$Er$_{x}$Cu$_2$O$_8$ 
as a matter of convenience.
For a reference, a superconducting sample
Bi$_2$Sr$_2$Ca$_{0.75}$Pr$_{0.25}$Cu$_2$O$_8$
(R=Ca$_{0.75}$Pr$_{0.25}$)
was also prepared.

\begin{figure}
 \begin{center}
  \includegraphics[width=8cm,clip]{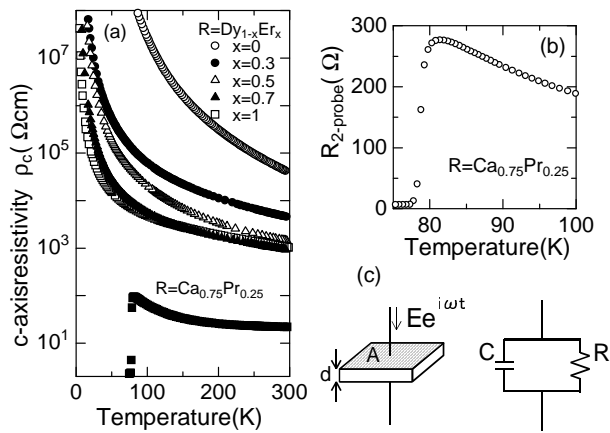}
 \end{center}
 \caption{
 (a) The out-of-plane resistivity 
 of Bi$_2$Sr$_2$RCu$_2$O$_8$ (R=Dy$_{1-x}$Er$_x$ and Ca$_{0.75}$Pr$_{0.25}$)
 measured using a two-probe technique,
 (b) The two-probe resistance ($R_{2-probe}$) for a superconducting sample
 Bi$_2$Sr$_2$Ca$_{0.75}$Pr$_{0.25}$Cu$_2$O$_8$,
 and (c) a schematic drawing for the complex impedance measurement.
 }\label{f1}
\end{figure}

Figure \ref{f1}(a) shows the temperature dependence of 
the out-of-plane resistivity for the measured samples.
Reflecting that the carrier concentration
increases with increasing Er content,
the magnitude systematically decreases with $x$.
For R=Ca$_{0.75}$Pr$_{0.25}$, 
the superconducting transition is observed slightly below 80~K.

We estimated the hole concentration per Cu ($p$) 
from the room-temperature thermopower \cite{Tallon},
and found $p$=0, 0.01, 0.015, 0.03, and 0.035 
for $x$=0, 0.3, 0.5, 0.7, and 1.0, respectively.
For R=Ca$_{0.75}$Pr$_{0.25}$,
$p$ was estimated to be 0.14 (underdoped region).
The uncertainty in $p$ was typically $\pm 0.005$ arising from
the scattered data in Ref.\cite{Tallon},
within which the estimated $p$ is consistent with 
the $p$ estimated from the Hall coefficient.

The complex impedance of the samples in the frequency
range of 10$^6$-10$^9$ Hz was measured
with a two-probe technique using an rf LCR meter (Agilent 4287A)
with a similar technique by B\"ohmer {\it et al}. \cite{Bohmer}
A low contact resistance was realized by uniformly 
painting the silver paste (Dupont 6838) 
on both sides of the $c$-plane surface,
followed by annealing at 873 K for 30 min. 
The contact resistance was 1-2 $\Omega$,
which can be safely neglected by comparison with 
the sample resistance (typically 100 $\Omega$).
In order to make the sample resistance high,
we cut the sample along the $a$ and $b$ directions as small as possible.
For the most conducting sample of R=Ca$_{0.75}$Pr$_{0.25}$,
the dimension was 0.2$\times$0.2$\times$0.05 mm$^3$.
Figure \ref{f1}(b) shows a typical two-probe resistance 
for R=Ca$_{0.75}$Pr$_{0.25}$.
Since the sample is superconducting, the resistivity rapidly
decreases below 80 K, and the residual resistance is 6.6 $\Omega$
that is the sum of the contact resistance and 
the probe resistance ($\sim 5\Omega$).
The sample resistance is 280 $\Omega$ at 80 K, which 
indicates the contribution of the contact resistance is less than 1 \%. 

As schematically shown in figure \ref{f1}(c), 
the measured impedance was understood as 
a parallel circuit consisting of a resistance $R(\omega)$ and 
a capacitance $C(\omega)$ which can be dependent on frequency $\omega=2\pi f$.
Then the complex impedance $Z$ is written as
\begin{equation}
 Z = \frac{R}{1+i\omega CR} = \frac{R}{1+(\omega CR)^2} 
  -i \frac{\omega CR^2}{1+(\omega CR)^2}
  \label{z}
\end{equation}
Since the capacitance is expressed in terms of the dielectric constant
$\varepsilon$ as $C=\varepsilon_0\varepsilon A/d$,
$\varepsilon$ is obtained from the complex impedance as
\begin{equation}
 \varepsilon = - \frac{d}{\omega\varepsilon_0 A} \frac{{\rm Im} Z}{|Z|^2}
\end{equation}
where $A$ and $d$ are the area and the distance of the capacitor,
and $\varepsilon_0$ is the dielectric constant of vacuum.

\begin{figure}
 \begin{center}
  \includegraphics[width=8cm,clip]{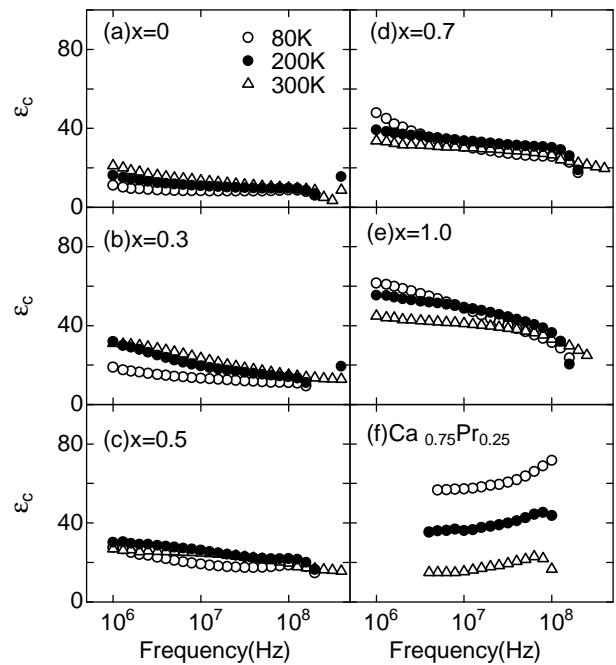}
 \end{center}
 \caption{
 The out-of-plane dielectric constants ($\varepsilon_c$) of 
 Bi$_2$Sr$_2$RCu$_2$O$_8$ (R=Dy$_{1-x}$Er$_x$ and Ca$_{0.75}$Pr$_{0.25}$).
 }\label{f2}
\end{figure}

%---------------------------------------------------------
%     Results and Discussion
%---------------------------------------------------------
\section{Results} 
Figure \ref{f2} shows the out-of-plane dielectric constant ($\varepsilon_c$). 
The magnitude of $\varepsilon_c$ reaches a fairly large value of 10-60.
Previously we found that the frequency dependence of 
$\varepsilon_c$ for $x=0$ is explained 
with the Debye model of dielectric relaxation \cite{Kitajima2,yanagi}.
Although the relaxation behaviour may not be clearly seen in figure \ref{f2}, 
$\varepsilon_c$ decreases with increasing frequency.
Sekhar {\it et al.} \cite{sekhar} measured $\varepsilon$ of
the polycrystalline samples of Bi$_2$Sr$_2$RCu$_2$O$_8$ (R=Sm and Y),
and found that $\varepsilon$ is as large as 10$^4$ at 300 K, 
and shows a relaxation behaviour.
Lunkenheimer {\it et al.} \cite{lunkenheimer} measured anisotropic 
dielectric constant of single-crystal La$_2$CuO$_4$,
and found the relaxation behaviour along the c-axis direction.
For other parent insulators, 
Shi \cite{shi} measured $\varepsilon$ of 
polycrystalline samples of (La,Gd)$_2$CuO$_4$, 
and Mazzara {\it et al.} \cite{mazzara} measured 
$\varepsilon_c$ of single-crystal PrBa$_2$Cu$_3$O$_7$.
These materials also show a large $\varepsilon$ with a relaxation behaviour,
and we conclude that this is generic 
in the parent insulator of HTSC.

It is quite difficult to accurately measure $\varepsilon_c$ for
a superconducting sample.
An upturn above 10$^7$ Hz for R=Ca$_{0.75}$Pr$_{0.25}$ is 
an artifact due to an inductance contribution
from the leads because of the small sample resistivity,
and an intrinsic dielectric response is likely to be independent of frequency.
The low-frequency data is also difficult to measure.
Equation (\ref{z}) requires that $\omega CR^2$
should be in the measurable range of the analyser 
(typically larger than 1$\Omega$)
but $\omega CR^2$ for R=Ca$_{0.75}$Pr$_{0.25}$ 
was too small near 10$^6$ Hz. 
A similar situation happened in the samples of $x$=0.7 and 1 
below 10$^6$ Hz, where we failed to measure $\varepsilon_c$.

An important finding is that $\varepsilon_c$ remains positive 
even in the superconducting sample of R=Ca$_{0.75}$Pr$_{0.25}$.
Kitano {\it et al.} \cite{Kitano} reported nearly 
the same $\varepsilon_c$ of Bi$_2$Sr$_2$CaCu$_2$O$_8$ 
(40 at 100 K at 10 GHz). 
This implies no divergence of $\varepsilon_c$ at IMT,
as Chen {\it et al.} \cite{Chen} suggested previously.
We further note that the out-of-plane charge response 
is not Drude-like, in the sense that $\varepsilon$ is 
negative below the plasma frequency in the Drude model.

\section{Discussion}
Let us discuss the magnitude and the frequency dependence of $\varepsilon_c$.
First, the large $\varepsilon_c$ is not due to phonons or lattice
instability, and should be attributed to an electronic excitation.
Unlike La$_2$CuO$_4$, Bi-2212 is chemically stable and shows 
no structure phase transition.
Second the electronic excitation should be in the CuO$_2$ plane.
We measured $\varepsilon_c$ for the layered cobaltate Bi-Sr-Co-O
that has a similar Bi$_2$O$_2$ layer,
and found a small $\varepsilon_c$ with essentially 
no relaxation behaviour \cite{terasaki}.
This indicates that neither  the other block layers nor the layered structure
can be an origin for the large $\varepsilon_c$.

Let us consider the electronic excitation 
through a simple Lorentz model written as 
\begin{equation}
 \varepsilon(\omega) = \varepsilon_{\infty} +
  \frac{f}{\omega_0^2-\omega^2 +i\gamma \omega} \label{lorentz}
\end{equation}
where $f$, $\omega_0$, $\gamma$ are the oscillator strength,
the resonance frequency, and the damping rate, respectively.
For an overdamped case $\gamma \gg \omega_0$,
Equation (\ref{lorentz}) reduces to the Debye model
of dielectric relaxation in the low frequency limit 
$\omega_0 \gg \omega$ as 
\begin{equation}
 \varepsilon(\omega) = \varepsilon_{\infty} +
  \frac{f}{\omega_0^2} \ \frac{1}{1+i\omega\tau}
  \label{debye}
\end{equation}
where $\tau = \gamma/\omega_0^2$.
Accordingly a large $\varepsilon$ implies
a small $\omega_0$ accompanied by a large $f$. 
We can demonstrate that $\varepsilon$ is small 
for a simple band insulator,
where $\omega_0$ and $f$ correspond to the band gap  
and the Drude weight of the valence electron.
Since these two are of the same order,
the second term of Equation (\ref{debye}) gives 
of the order of unity for $\omega\to 0$.
Note that $\varepsilon$ is also small for a Mott insulator,
because the Mott-Hubbard gap is of the same order as the band gap.
In other words, a single-particle gap is too large to
give a large $\varepsilon$.

A collective-excitation gap is most likely to be
responsible for a large $\varepsilon$. 
It is known that the charge-density-wave (or charge-ordered) 
materials, such as K$_{0.3}$MoO$_3$ \cite{Cava},
LuFe$_2$O$_4$ \cite{ikeda}, (Pr,Ca)MnO$_3$ \cite{arima}
exhibit a large $\varepsilon$ with relaxation behaviour,
where $\omega_0$ is the pinning frequency of the charge density.
This implies that a charge density modulation or a charge fluctuation 
exists in the CuO$_2$ plane,
which in principle can occur in the doped antiferromagnet.
Unlike the case of a doped semiconductor,
holes in the parent insulator do not distribute uniformly.
They tend to condense in order that 
they should minimise the exchange-energy loss of
the antiferromagnetic background.
Prime examples are the charge stripe \cite{stripe}
and the phase separation \cite{maekawa}. 
Thus we think that the observed dielectric response 
is a piece of evidence for the charge modulation in the parent insulator.

Recent scanning tunnel microscope/spectroscopy 
has revealed unexpected inhomogeneous electronic states of HTSC. 
Pan {\it et al.} \cite{pan} found that the local density of states
are inhomogeneous in the CuO$_2$ plane of
the optimally doped Bi-2212,
where the doped carriers form a metallic patch.
Very recently Howald {\it et al.} \cite{howald} also found
superconducting and non-superconducting phases (at a nanometer scale)
coexist in the CuO$_2$ plane of the underdoped Bi-2212.
This type of charge inhomogeneity could cause large $\varepsilon$,
and is consistent with our observations.

\begin{figure}
\begin{center}
 \includegraphics[width=8cm,clip]{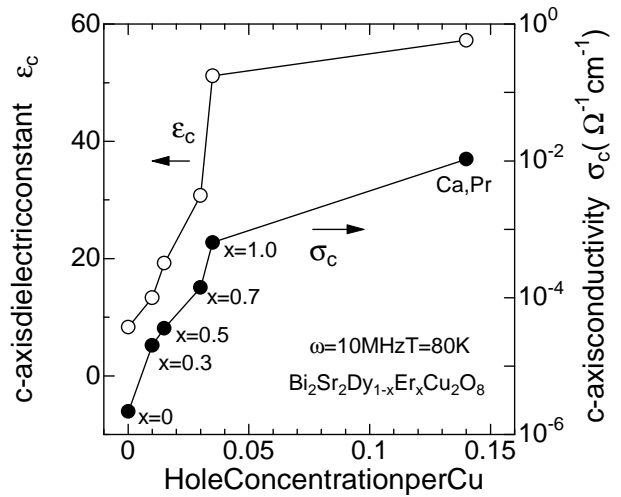}
\end{center}
 \caption{
 The out-of-plane dielectric constant and conductivity
 Bi$_2$Sr$_2$RCu$_2$O$_8$ at 10~MHz 
 plotted as a function of hole concentration per Cu ($p$). 
 The $p$ was estimated by measuring the room-temperature thermopower. 
 }\label{f3}
\end{figure}

Next let us discuss the the carrier concentration dependence.
In figure \ref{f3}, $\varepsilon_c$ and 
the out-of-plane conductivity $\sigma_c$ 
for a representative frequency of 10$^7$ Hz at 80 K
are plotted as a function of $p$.
Most importantly, finite values of $\varepsilon_c$ and $\sigma_c$ 
are observed simultaneously.
This contrasts strikingly with a case of a doped semiconductor,
in which the insulating region is characterised by 
a finite $\varepsilon$ and $\sigma=0$  \cite{castner},
whereas the metallic region is characterised by 
a finite $\sigma$ and $\varepsilon=0$ \cite{rosenbaum}.
The $p$ dependence in figure \ref{f3} suggests that
the parent insulator is a mixture of an insulating and metallic phases.
The increase of $\sigma_c$ with $p$ suggests that
carrier-doping makes locally metallic patches
in the insulating background.
The patches cause a large $\varepsilon_c$ by fluctuating 
its position and a finite $\sigma_c$ 
through the proximity effects. 
Then the charge dynamics can be understood in terms of 
a percolative mixture of the insulating and metallic phases.
Accordingly  the insulator-superconductor transition in Bi-2212 
occurs, when the metallic (superconducting) patches 
are connected with one another over a whole sample with proximity effects.

Now we will consider the anisotropy of $\varepsilon$.
Based on the inhomogeneity scenario, $\varepsilon$ is
enhanced to be $\varepsilon_{\infty} + f/\omega_0^2$ 
for $\omega \to 0$, where $f$ is the Drude weight of the metallic 
patch that is inversely proportional to the effective mass.
It is therefore expected that $\varepsilon_{ab}$ is 
larger by the effective-mass ratio ($10^4$) than $\varepsilon_c$,
whereas the frequency and temperature dependences are nearly identical.
This is what we observed in the preliminary measurements for
$\varepsilon_{ab}$ of R=Dy and Er \cite{yanagi}.
In a superconducting sample, 
the metallic patches are connected from edge to edge,
and the carriers can move freely along the in-plane direction.
For Bi-2212, carriers near the cold spot ($\pi/2, \pi/2$)
dominates the in-plane transport \cite{millis}, 
and are little affected by the inhomogeneity,
because the local density of states is rather homogeneous along 
the ($\pi/2, \pi/2$) (nodal) direction \cite{pan}.
This means that a Drude-like charge response 
can occur along the in-plane direction. 
Then $\varepsilon_{ab}$ changes with doping from positive to negative across the
IMT, at which the divergence of $\varepsilon_{ab}$ would be seen.
On the other hand, owing to the zero matrix element \cite{andersen},
carriers near ($\pi/2, \pi/2$) cannot move
along the out-of-plane direction,
and consequently cannot screen the inhomogeneity in the CuO$_2$ plane.
This is the reason the dielectric response along the out-of-plane direction
remains unchanged at IMT.

Finally we will compare our results with those by 
Chen {\it et al.} \cite{Chen}
At a qualitative level, our observations are similar to theirs:
(1) $\varepsilon_c$ increases with decreasing frequency,
and (2) $\varepsilon_c$ does not diverge at IMT. 
However our measurement are free from the lattice instability,
and includes the data for a superconductor, 
from which we have successfully shown 
that the items (1)(2) are generic for HTSC.
Another discrepancy is that 
we understood the frequency dependence as the dielectric relaxation,
while they employed the variable range hopping.
Considering $\epsilon$ of other parent insulators,
we think that the relaxation behaviour is generic.
Perhaps this comes from the different temperatures measured.
They discussed $\varepsilon$ at 4 K,
where the resistivity is also described in terms of variable range hopping.
We found that the resistivity anisotropy is strongly dependent
on temperature \cite{Kitajima}.
The resistivity is almost isotropic at 4 K, 
and the parent insulator behaves three-dimensional.
On the contrary, the resistivity is largely anisotropic above 80 K, 
and in particular the ``confinement'' is observed even in the parent insulator.
Thus it would be natural that the frequency dependence of $\varepsilon$
is different between theirs and ours.

%---------------------------------------------------------
%    Summary
%---------------------------------------------------------
\section{Summary}
We prepared single crystals of Bi$_2$Sr$_2$Dy$_{1-x}$Er$_x$Cu$_2$O$_8$,
and measured the out-of-plane dielectric constants 
in the temperature range of 80-300 K 
and in the frequency range of $10^6$-$10^9$ Hz. 
The dielectric constant of the parent insulator 
is characterised by the large magnitude and relaxation behaviour, which
would come from the charge inhomogeneity recently observed in
the scanning tunnel microscope experiments.
The doping dependence of the dielectric constant and
the conductivity suggests that
the charge dynamics is understood as a percolation 
of the insulating and metallic phases,
and that the insulator-superconductor transition happens 
when the metallic (superconducting) patches 
are connected with one another over a whole sample.

%---------------------------------------------------------
%   Acknowledgements
%---------------------------------------------------------
\section*{Acknowledgements}

The authors would like to thank S Tajima, K Yoshida and S Tanaka 
for the technical support of rf impedance measurement. 
They also appreciate T Sugaya and T Mizuno 
for the preparation of the single crystals 
and the measurement of the thermopower,
and T Kitajima for collaboration at the early stage of the present study.
This work is partially supported by 
The Kawakami Memorial Foundation, and 
Waseda University Grant for Special Research Projects (99A-556). 

%---------------------------------------------------------
%    References
%---------------------------------------------------------

\end{document}